\documentclass{elsart}
\usepackage{natbib}
\usepackage{psfig}
\begin{document}
\runauthor{De~Breuck \etal}
\begin{frontmatter}
\title{CO Emission from $z>3$ Radio Galaxies}
\author[iap]{Carlos De Breuck}
\author[iram]{Roberto Neri}
\author[iap]{Alain Omont}

\address[iap]{Institut d'Astrophysique de Paris, 98bis Boulevard Arago, 75014 Paris, France}
\address[iram]{IRAM, 300 Rue de la Piscine, 38406 St. Martin-d'H\`eres, France}
\begin{abstract}
We report on the detection of the CO(4$-$3) line with the IRAM Plateau de Bure Interferometer in two $z>3$ radio galaxies, doubling the number of successful detections in such objects. A comparison of the CO and Ly$\alpha$ velocity profiles indicates that in at least half of the cases, the CO is coincident in velocity with associated HI absorption seen against the Ly$\alpha$ emission. This strongly suggests that the CO and HI originate from the same gas reservoir, and could explain the observed redshift differences between the optical narrow emission lines and the CO. The CO emission traces a mass of H$_2$ 100$-$1000 times larger than the HI and HII mass traced by Ly$\alpha$, providing sufficient gas to supply the massive starbursts suggested by their strong thermal dust emission.
\end{abstract}
\begin{keyword}
radio lines: galaxies - galaxies: active - galaxies: formation
\end{keyword}
\end{frontmatter}

\section{Introduction}
High redshift radio galaxies (HzRGs; $z>2$) are often argued to be identified with massive forming ellipticals undergoing a major starburst. There are several arguments for ongoing star formation in HzRGs, including (i) the detection of stellar photospheric absorption lines seen in deep optical spectroscopy (Dey \etal\ 1997; De~Breuck \etal, in prep.) and (ii) their strong thermal emission at sub-mm wavelengths, especially at $z>3$ (Archibald \etal\ 2001; Reuland \etal, these proceedings). This thermal dust emission can be heated by massive star formation, but direct heating by the AGN is also likely to contribute (e.g.\ Sanders \& Mirabel 1996; Omont \etal\ 2001). One possibility to distinguish between both mechanisms is to search for CO emission tracing the massive gas reservoir needed to feed the massive starbursts. Such spatially resolved CO emission has indeed been detected around a HzRG (Papadopoulos \etal\ 2000) and quasars up to $z=4.7$ (Omont \etal\ 1996).

Despite intensive search campaigns (Evans \etal\ 1996; van Ojik \etal\ 1997b), the CO lines in HzRGs have long defied detection. The pre-selection of objects with strong thermal dust emission detected with SCUBA at 850~$\mu$m, and the use of the sensitive IRAM Plateau de Bure Interferometer have lead to the first detections of the CO (4$-$3) line in two $z>3$ radio galaxies (Papadopoulos \etal\ 2000). 
These detections are expected, given the presence of large quantities of ionized and neutral hydrogen around many HzRGs. The ionized gas is seen in the giant Ly$\alpha$ halos (see the contributions by van Breugel and Villar-Mart\'\i n), while the surrounding neutral gas often causes associated HI absorption in the Ly$\alpha$ lines of HzRGs (e.g. van Ojik \etal\ 1997a; Jarvis \etal\ 2003).
However, no direct relationship between these different gas components has yet been found. In this paper, we report on the detection of CO(4$-$3) in two additional HzRGs, which are coincident with the associated HI absorption in the Ly$\alpha$ lines.

\begin{figure*}
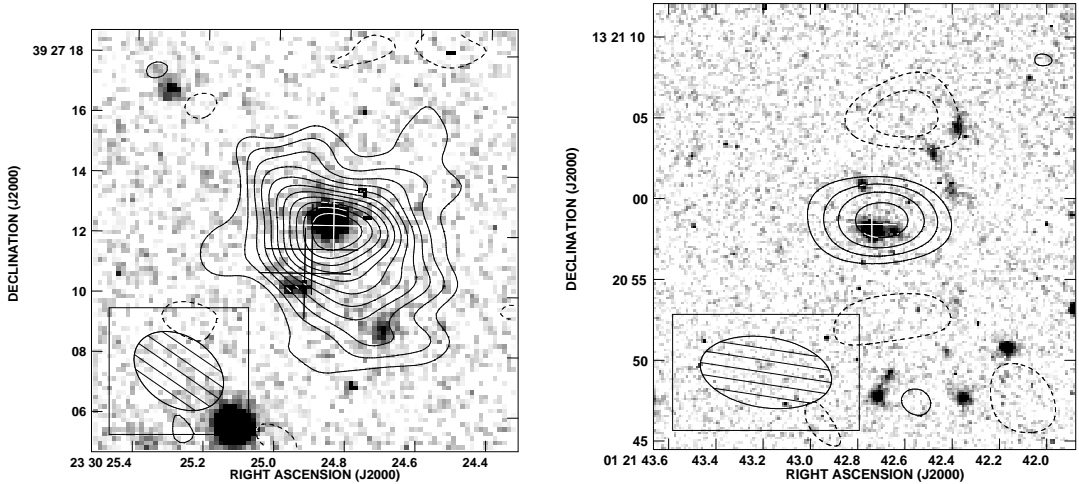

\begin{tabular}{c c}
\psfig{file=B3K.COINT.PS,angle=-90,width=7cm}&
\psfig{file=TNJ0121K.CO43.ZOOM.PS,angle=-90,width=7cm}\\
\end{tabular}
\caption{Integrated CO~$J=4-3$ emission overlaid on Keck/NIRC $K-$band images for B3~J2330+3927 ($z=3.087$; left) and TN~J0121+1320 ($z=3.517$; right). Contour levels start at 2~$\sigma$, and increase by 1~$\sigma$, with $\sigma=0.2$~mJy~Beam$^{-1}$ and $\sigma=0.3$~mJy~Beam$^{-1}$, respectively (negative contours are dotted). The crosses mark the positions of the radio components. The FWHM of the synthesized beam is shown in the bottom left corner.}
\label{KCOoverlays}
\end{figure*}

\section{A Search for CO in HzRGs using IRAM Plateau de Bure}

During the past decade, CO lines have been detected in more than a dozen high redshift objects, primarily quasars out to $z=4.7$ (see Cox, Omont \& Bertoldi 2002 for a recent review). Most of these detections have been made with the IRAM Plateau de Bure Interferometer (PdBI), which is currently the most sensitive instrument at mm wavelengths. We have started a renewed search for CO lines in HzRGs with the PdBI to complement the quasar observations, which do not allow a detailed comparison with the optical properties due to the dominating AGN emission. 

Our targets are selected from the SCUBA survey of $z>3$ radio galaxies (Reuland, these proceedings), retaining only sources with $S_{\rm 850 \mu m} > 5$~mJy and a $>5\sigma$ detection. In order to pick up spatially extended emission, we use the most compact array configuration. The total bandwidth is 560~MHz, which, given the high accuracy of the optical redshifts determined from the non-resonant HeII~$\lambda$1640\AA\ line, should be sufficient to pick up the often broad ($>$500 km~s$^{-1}$) CO(4$-$3) lines. However, after an initial analysis of the first data, we often found an offset between the optical and CO redshifts (see Table~1), and shifted the central frequency to better cover the entire line width. We return to this difference in the next section. For further details on the observations, see De~Breuck \etal\ (2003).

The first results from our survey have been very promising: we successfully detected the CO(4$-$3) line in B3~J2330+3927 (De~Breuck \etal\ 2003) and TN~J0121+1320. Figure~\ref{KCOoverlays} shows the integrated CO emission, overlaid on Keck/NIRC $K-$band images, and Table~\ref{obsparms} lists the observational parameters. In TN~J0121+1320, the emission appears unresolved, but in B3~J2330+3927, the emission appears spatially extended towards a nearby object in the $K-$band image. Additional observations are be needed to confirm this extent.

\begin{table*}
\caption{Observational Parameters of $z>3$ Radio Galaxies with CO Detections.}
\label{obsparms} 

\begin{center}
\begin{tabular}{l l l l l c c}
\hline 
Source & $z_{\rm opt}$ & $z_{\rm CO}$ & $\Delta V_{\rm CO}$ & $S_{\rm CO}\Delta V$ & Ref. spec. & Ref. CO \\
  & & & km s$^{-1}$ & Jy km s$^{-1}$ & & \\
\hline 
B3~J2330+3927 & 3.087  & 3.094 & 500 & 1.3$\pm$0.3 & a & a \\
TN~J0121+1320 & 3.516  & 3.520 & 700 & 1.2$\pm$0.4 & b & c \\
6C~J1908+7220 & 3.5356 & 3.532 & 530 & 1.62$\pm$03 & b & d \\
4C~60.07      & 3.788  & 3.791 & $>$1000 & 2.5$\pm$0.43 & e & d \\
\hline 
\end{tabular}
\end{center}
References: (a) De~Breuck \etal\ (2003); (b) De~Breuck \etal\ (2001); (c) this paper; (d) Papadopoulos \etal\ (2000); (e) R\"ottgering \etal\ (1997).
\end{table*}

\section{Comparison of CO and Ly$\alpha$ velocity profiles}

Table~\ref{obsparms} indicates that, despite uncertainties due to the low S/N of the CO detections, there is often a significant offset between the redshift of the optical narrow emission lines and the CO lines. Such offsets are also seen in high S/N CO detections of several high redshift quasars (Guilloteau \etal\ 1999; Cox \etal\ 2002a).

\begin{figure}
\psfig{file=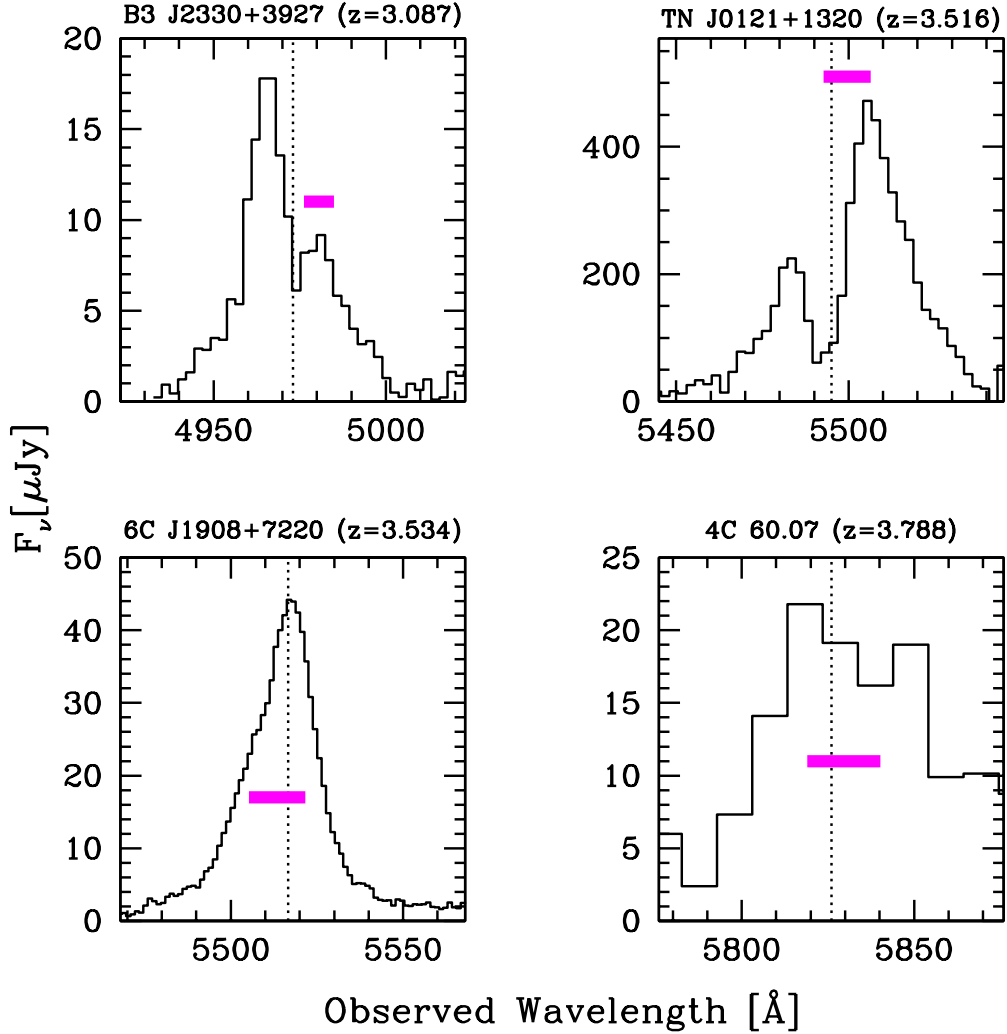,width=14cm}
\caption{Ly$\alpha$ velocity profiles of the 4 HzRGs with detected CO(4$-$3) lines. The horizontal bar indicates the full width of the CO emission, and the dotted vertical line the systemic redshift, as determined from the non-resonant HeII~$\lambda$1640\AA\ line.}
\label{lyaprofiles}
\end{figure}

Figure~\ref{lyaprofiles} compares the velocity profiles of the Ly$\alpha$ lines with the full velocity width of the CO lines from Papadopoulos \etal\ (2000) and our observations. In both B3~J2330+3927 and TN~J0121+1320, the CO lines correspond very closely in velocity space to the associated HI absorbers. 6C~J1908+7220 is a rare HzRG with broad absorption lines (BAL) and an asymmetrical Ly$\alpha$ profile, both spatially and in velocity (Dey 1999). Because the BAL absorption occurs very near the AGN (e.g. Weymann 1991), there is no 'background Ly$\alpha$ source' against which we could detect large-scale associated HI absorption. The spectrum of 4C~60.07 has very low resolution and S/N and does not provide reliable information on the velocity profile.
We conclude that in at least half of the HzRGs, the CO emission appears coincident with associated HI absorption. This strongly suggests that both the CO (which traces the H$_2$) and HI originate from the same gas reservoir surrounding the host galaxy. 

High resolution spectroscopy of the Ly$\alpha$ and CIV~$\lambda$1549\AA\ lines in two HzRGs (Binette \etal\ 2000; Jarvis \etal\ 2003; Wilman, these proceedings) has shown that the absorbers cannot be co-spatial with the emitting gas, but should be located in a surrounding shell. The spatially resolved CO emission may thus be originating from such shells. The presence of CO emission implies that these reservoirs cannot be primordial, which is consistent with the super-solar metalicity seen in the absorber of 0200+015 (Jarvis \etal\ 2003). 
Baker \etal\ (2002) argue that dust is mixed with the CIV-absorbing gas in radio-loud quasars, also suggesting that these reservoirs consist of enriched material. This material may have been deposited by previous mergers or starburst-driven superwinds (e.g. Taniguchi \etal\ 2001, Furlanetto \& Loeb 2003).

\section{Mass estimates}

It is of interest to compare the mass estimates from the CO emission, and the Ly$\alpha$ emission and absorption. Such estimates are very uncertain because we need to make a large number assumptions on the physical state and geometrical distribution of the gas (see De Breuck \etal\ 2003 for a discussion). Nevertheless, in the case of B3~J2330+3927, it is clear that the CO traces a 100 to 1000 larger mass of H$_2$ than the HI and HII derived from Ly$\alpha$. 
If both the CO and Ly$\alpha$ originate from the same gas reservoir, as suggested by the velocity profiles in Fig.~\ref{lyaprofiles}, this implies that CO is a much better tracer of the gas mass than Ly$\alpha$. High resolution CO maps of HzRGs will therefore provide a more complete picture of the formation stages of massive galaxies. However, to pick up all components, one needs to be sure to cover the entire velocity width of the CO emission, which can be offset from the optical emission lines by $>$500~km~s$^{-1}$ (see e.g. Fig.~1 of Papadopoulos \etal\ 2000). This presently requires multiple setups with the PdBI, but it will be straightforward with the new 4~GHz wide receivers, scheduled to be commissioned in 2004.

{\it Acknowledgments:}

IRAM is supported by INSU/CNRS (France), MPG (Germany) and IGN (Spain).
This work was supported by a Marie Curie Fellowship of the European Community programme 'Improving Human Research Potential and the Socio-Economic Knowledge Base' under contract number HPMF-CT-2000-00721.

\end{document}